\documentclass[%
 reprint,
 amsmath,amssymb,
 aps,
]{revtex4-2}

\usepackage{graphicx}
\usepackage{dcolumn}
\usepackage{bm}
\usepackage{color,soul}
\usepackage{amsmath}
\usepackage{xcolor}

\begin{document}


\title{Dynamic hybrid metasurfaces}

\author{Sajjad Abdollahramezani$^{1}$}
\author{Omid Hemmatyar$^{1}$}
\author{Mohammad Taghinejad$^{1}$}
\author{Hossein Taghinejad$^{1}$}
\author{Yashar Kiarashinejad$^{1}$}
\author{Mohammadreza Zandehshahvar$^{1}$}
\author{Tianren Fan$^{1}$}
\author{Sanchit Deshmukh$^{2}$}
\author{Ali A. Eftekhar$^{1}$}
\author{Wenshan Cai$^{1,3}$}
\author{Eric Pop$^{2}$}
\author{Mostafa El-Sayed$^{4}$}
\author{Ali Adibi$^{1,}$}

\email{ali.adibi@ece.gatech.edu}

\affiliation{$^{1}$School of Electrical and Computer Engineering, Georgia Institute of Technology, 778 Atlantic Drive NW, Atlanta, Georgia 30332-0250, United States}
\affiliation{$^{2}$Department of Electrical Engineering, Department of Materials Science and Engineering, Precourt Institute for Energy, Stanford University, Stanford, California 94305, United States}
\affiliation{$^{3}$School of Materials Science and Engineering, Georgia Institute of Technology, 801 Ferst Drive NW, Atlanta, Georgia 30332-0295, United States}
\affiliation{$^{4}$Laser Dynamics Laboratory, School of Chemistry and Biochemistry, Georgia Institute of Technology, Atlanta, Georgia, 30332-0400, United States}

\date{\today}



\begin{abstract}

Efficient hybrid plasmonic-photonic metasurfaces that simultaneously take advantage of the potential of both pure metallic and all-dielectric nanoantennas are identified as an emerging technology in flat optics. Nevertheless, post-fabrication tunable hybrid metasurfaces are still elusive. Here, we present a reconfigurable hybrid metasurface platform by incorporating the phase-change material Ge$_{2}$Sb$_{2}$Te$_{5}$ (GST) into metal-dielectric meta-atoms for active and non-volatile tuning of properties of light. We systematically design a reduced-dimension meta-atom, which selectively controls the fundamental hybrid plasmonic-photonic resonances of the metasurface via the dynamic change of optical constants of GST without compromising the scattering efficiency. As a proof-of-concept, we experimentally demonstrate miniaturized tunable metasurfaces that control the amplitude and phase of incident light necessary for high-contrast optical switching and anomalous to specular beam deflection, respectively. Finally, we leverage a deep learning-based approach to present an intuitive low-dimensional visualization of the enhanced range of response reconfiguration enabled by the addition of GST. Our findings further substantiate dynamically tunable hybrid metasurfaces as promising candidates for the development of small-footprint energy harvesting, imaging, and optical signal processing devices.

\end{abstract}


\maketitle


\section{Introduction}

Optical metasurfaces, formed by an array of subwavelength-spaced nanoscatterers (or meta-atoms)  on a planar interface, have garnered ever-increasing attention lately \cite{yu2011light,kuznetsov2016optically}. Plasmonic metasurfaces, with strong field enhancement near their meta-atoms, offer good scattering directionality and strong light-matter interaction \cite{schuller2010plasmonics,ding2018review}. However, they suffer from nonradiative losses, low-quality-factor (low-Q) modes, limited scattering cross-section, low damage-threshold, and CMOS-incompatibility. As a viable alternative, low-loss all-dielectric meta-atoms can effectively scatter the incident light, leading to exceptionally directional forward and backward scatterings \cite{staude2013tailoring, arbabi2015dielectric,jahani2016all, zhan2016low, kamali2018review, taghinejad2019all, hemmatyar2019full, abdollahramezani2020meta} at the cost of lower field confinements and larger mode volumes, which hamper deep-subwavelength efficient meta-atoms. In this regard, employing hybrid metal-dielectric meta-atoms to fully take advantage of highly directional radiation patterns while preserving high coupling efficiency and moderate dissipation loss is promising for flat optics. Thus far, hybrid architectures composed of high-index nanoscatterers as the director component in combination with resonant plasmonic nanoantennas as the feed element or non-resonant plasmonic back-reflectors as the mode enhancer have been demonstrated \cite{oulton2008hybrid,metzger2014doubling,guo2016multipolar,yang2017low,lepeshov2019hybrid}.

Despite the usefulness of static metasurfaces, major efforts are still needed to realize highly integrated, energy-efficient, easy-to-fabricate, and low-cost reconfigurable meta-atoms that can revolutionize modern flat optics technology to surpass conventional spatial light modulators based on digital micromirrors or liquid crystal (LC) cells. A series of materials and architectures with external stimuli have been used for active tuning of meta-devices \cite{hail2019optical,shaltout2019spatiotemporal}. Tuning the refractive index of semiconductors using the thermo-optic effect \cite{wang2020fast, zhang2018ultra}, controlling the electro-optic properties of transparent conductive oxides utilizing free carrier effects \cite{kafaie2018dual,shirmanesh2019electro}, manipulating the effective physical state of LC based on phase transition effects \cite{komar2017electrically,li2019phase}, and adjusting the stretch of elastomeric substrates using strain effects \cite{kamali2016highly} have been the mainstream of recent researches. Despite interesting results, most proposed techniques suffer from large reconfigurable unit cells, limited operation bandwidth, low reconfiguration speed, and complex fabrication processes, to name a few. A key challenge in such structures is the need for a fast tunable material with large control over its optical properties (e.g., change in the index of refraction $\Delta$n$~\approx~1$). This becomes even more essential when we consider the low-Q nature of existing static metasurfaces. Phase-change materials (PCMs) with unprecedented non-volatile change in their index of refraction upon transition between amorphous and crystalline states, can be a reliable resolution to this challenge \cite{wuttig2007phase,zhang2019broadband,delaney2020new}. Integrating PCMs with hybrid metal-dielectric metasurfaces with inherently moderate-Q nanoresonators can further enhance the dynamic range and efficiency of the reconfiguration process.

\begin{figure*}
	\centering
	\includegraphics[trim={0cm 0cm 0cm 0cm},width=0.93\textwidth, clip]{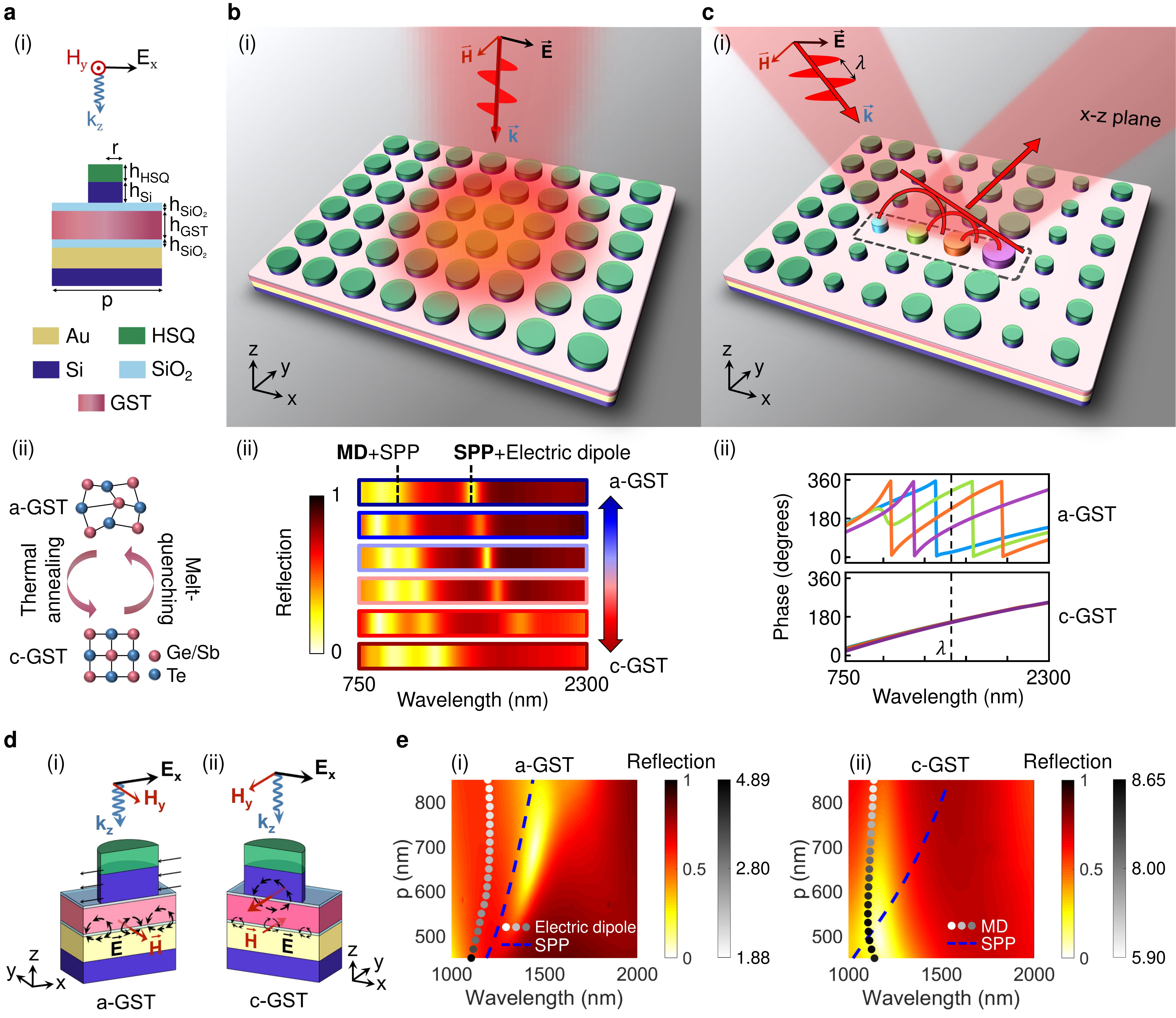}
	\caption{\textbf{Concept images and working principle of the reconfigurable GST-assisted hybrid metal-dielectric metasurface}.
	\textbf{(a)} Detailed cross sectional view of a meta-atom consisting of a Si nanodisk on a GST film laid on top of an optically-thick Au substrate (i), and the generic scheme of atomic distribution of amorphous/crystalline GST (a-GST/c-GST) before/after thermal annealing (ii).
	\textbf{(b)} Perspective view of a meta-switch illuminated by a broadband near-IR light (i), and the evolution of fundamental plasmonic-photonic modes upon transition of the GST state from amorphous to crystalline in multiple intermediate levels (ii). Texts with the bold font (i.e., \textbf{MD} and \textbf{SPP}) represent the dominant mode in each resonance. 
	\textbf{(c)} (i) Schematic representation of a meta-deflector illuminated with a laser light, and (ii) the spectral response of each color-coded Si nanodisk sketched in the separated meta-molecule in (i) for a-GST and c-GST cases. Each meta-atom is responsible for $90^{\circ}$ and $0^{\circ}$ phase shift increment when GST is in its amorphous and crystalline state, respectively.
	\textbf{(d)} Conceptual representation of the excited strong SPP mode that interacts with the electric dipole mode in the case of a-GST (i), and the enhanced MD that interferes with the SPP mode in the case of c-GST (ii). The dominant and secondary components are shown with thick and thin arrows, respectively.
	\textbf{(e)} Two-dimensional (2D) reflection maps for the meta-atom in \textbf{(a)} with (i) amorphous and (ii) crystalline GST states as functions of period ($p$) and wavelength ($\lambda$). The gray-scale-coded dotted lines in (i) and (ii) represent the electric field enhancement induced by the electric dipole resonance at the gap center between two neighboring meta-atoms, and the magnetic field enhancement induced by the MD mode at the center of the pseudo-pillar (formed by the Si nanodisk and the subjacent GST segment), respectively. Gray color bars in both (i) and (ii) display the field amplitudes normalized to the incident plane wave. The evolution of the SPP mode for both amorphous and crystalline GST states is displayed by the blue dashed line. The structural design parameters are chosen as $h_{\textrm{HSQ}}$ = 90 nm, $h_{\textrm{Si}}$ = 100 nm, $h_{\textrm{SiO}_{2}}$ = 5 nm, $h_{\textrm{GST}}$ = 70 nm, while for \textbf{(b} (i)\textbf{)} $r$ = 280 nm, $p$ =650 nm, \textbf{(c} (i)\textbf{)} $r$ = [110 (blue), 155 (green), 185 (orange), 265 (purple)] nm, $p$ = 650 nm, and \textbf{(e)} $r$ = 200 nm, $p$ varies according to \textbf{(a} (i)\textbf{)}. $\vec{\textrm{E}}$, $\vec{\textrm{H}}$, and $\vec{\textrm{k}}$ are the electric field, magnetic field, and Poynting vector, respectively.}
	\label{fig_1}
\end{figure*}

Germanium antimony telluride (Ge$_{2}$Sb$_{2}$Te$_{5}$ or GST for short), with ultrafast switching speeds (10's to 100's of nanoseconds), considerable spatial control and miniaturization (down to nm sizes), high switching robustness (up to $10^{15}$ cycles), good thermal stability (up to several 100 $^{\circ}$C), and CMOS compatibility, has been vastly exploited in the commercial rewritable optical disk storage and resistive-switching electronic memory technologies \cite{wang2016optically,michel2014reversible,michel2019advanced,taghinejad2020ito}. The non-volatile nature of optical/electrical changes in GST makes it superior in terms of static energy consumption. Moreover, the intermediate phase transition of GST (between amorphous and fully-crystalline phases), empowers meta-atoms to induce arbitrary transformations in amplitude, phase, and even polarization of the incident optical wavefront, leading to reprogrammable pixelated multifunctional metasurfaces \cite{wuttig2017phase,ding2019dynamic,abdollahramezani2020tunable, shalaginov2020design}. Recently, reconfigurable nanophotonic devices based on hybrid GST-metal meta-atoms have been proposed \cite{gholipour2013all, yin2015active}. Nevertheless, the optical loss of such platforms hinders many real-world applications including lenses and phase plates. Although patterned PCM-dielectric metasurfaces are promising platforms, achieving truly reversible subwavelength pixels offering full 360$^{\circ}$ optical phase coverage with high performance is still challenging due to low-Q (and relatively weak light-matter interaction) even in the mid-infrared (mid-IR) spectral range \cite{tian2019active,de2020reconfigurable,leitis2020all}.

In this paper, we leverage the unique features of GST to demonstrate the active control of the scattering properties of a hybrid metal-dielectric metasurface. For the first time, to our knowledge, we experimentally demonstrate a reconfigurable architecture with non-volatile engineering of the characteristic resonances of the magnetic Mie (or magnetic dipole (MD)) mode and the electric dipole mode supported by the engineered dielectric meta-atoms, and the surface plasmon polariton (SPP) mode supported by a plasmonic back reflector. This architecture enables rich hybrid plasmonic-photonic modes without imposing fabrication complexities. Full-wave simulation results reveal the enhancement of light-matter interaction within the GST layer for on-demand generation/cancellation and strengthening/weakening of such modes at specific wavelengths. As a proof-of concept, we demonstrate two classes of dynamic metadevices: i) a polarization-insensitive absorptive metasurface (or a meta-switch) exhibiting high absorption performance with a characteristic modulation depth comparable to the alternative state-of-the-art devices, and ii) a beam-steering metasurface (or a meta-deflector) for deflecting a monochromatic light beam to the anomalous/specular angle upon the active transition of the GST state. The experimental results of this paper suggest the potential of the demonstrated hybrid architecture for realization of a wide range of multifunctional meta-devices for applications like imaging, sensing, computing, and ranging, just to name a few.

\section{Results}

\subsection{Metasurface working principle}

Schematic representations of the proposed tunable GST-assisted hybrid metal-dielectric meta-atom is shown in Fig.~\ref{fig_1}a (i). It consists of a silicon (Si) nanodisk sitting on top of a GST layer deposited on an optically-opaque gold (Au) back reflector. In the near-IR regime, GST essentially serves as a functional dielectric medium whose optical properties can be finely tuned by applying an external stimulus \cite{rios2015integrated,wang2016optically}. This leads to distinct non-volatile intermediate states enabling step-wise manipulation of the optical wavefront even at a pixel level. Complex refractive index of GST in the two extreme states, i.e., amorphous (a-GST associated with ${0\%}$ crystallization level, i.e., L$^\textrm{0\%}$) and crystalline (c-GST associated with ${100\%}$ crystallization level, i.e., L$^\textrm{100\%}$), and 4 intermediate levels are demonstrated in the Supplementary Fig.~1. The optical constants of GST in the intermediate levels are calculated using a well-known effective medium theory (see Methods for more details). Upon thermal-based conversion to c-GST on top of a hotplate, a remarkable contrast is induced in the complex index of a-GST (e.g., $\Delta$n$~\approx 2.5$ and $\Delta$k$~\approx 0.9$ at $\lambda = 1550$ nm, with n and k being the real and imaginary parts of the index of refraction) due to a pronounced change in the alloy chemical formation (from the covalent bonding in a-GST to the resonance bonding with high electronic polarizability in c-GST \cite{shportko2008resonant}) as conceptually shown in Fig.~\ref{fig_1}a (ii). 

Perspective views of two classes of reprogrammable metasurfaces, i.e., a meta-switch and a meta-deflector, that employ uniform matrices of hybrid meta-atoms and meta-molecules are depicted in Figs. \ref{fig_1}b (i) and \ref{fig_1}c (i), respectively. Ultimate manipulation of the amplitude and phase properties of reflected light from these meta-devices is enabled through gradual and abrupt transition of underlying hybrid modes thanks to the dynamically tunable GST layer (see Figs. \ref{fig_1}b (ii) and \ref{fig_1}c (ii)). The representative flow diagram of the fabrication process of such meta-devices is shown in Supplementary Fig.~2, and its detailed explanation is provided in Methods.

As conceptually represented in Fig.~\ref{fig_1}d, the resonance mode of the meta-atom in Fig.~\ref{fig_1}a (i) is formed by interaction of the plasmonic and photonic modes in both cases of a-GST and c-GST. This interaction enables truly-subwavelength vertical confinement and horizontal enhancement of the electromagnetic field, respectively. It also facilitates on-demand high radiation at the low-loss dielectric-air interface with more efficiency than conventional pure plasmonic or all-dielectric configurations. In order to reveal the underlying nature of light-matter interaction at different crystallization states of GST in the hybrid meta-atom in Fig.~\ref{fig_1}a (i), we conduct full-wave numerical simulations to study the mode profiles of meta-atoms and reflection properties of the metasurfaces for all relevant cases. The details of simulations are described in Methods.

\begin{figure*}
	\centering
	\includegraphics[trim={0cm 0cm 0cm 0cm},width=0.85\textwidth, clip]{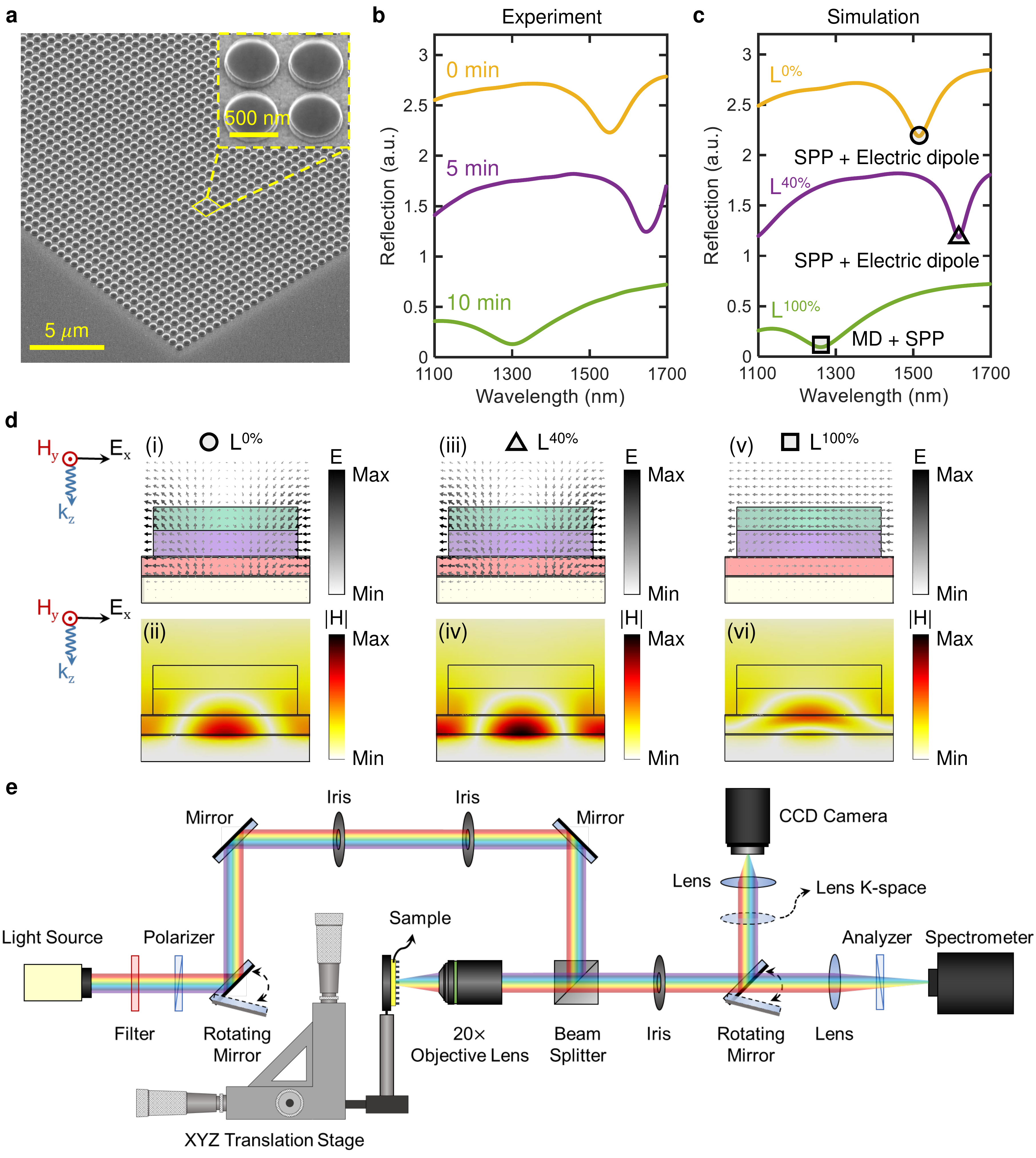}
	\caption{\textbf{Experimental realization of the dynamic meta-switch}. \textbf{(a)} Tilted SEM of the as-fabricated tunable meta-switch. The structural parameters are $p$ = 650 nm, $r$ = 280 nm, $h_\textrm{Au}$ = 100 nm, $h_\textrm{{SiO2}}$ = 5 nm, $h_\textrm{GST}$ = 75 nm, $h_\textrm{Si}$ = 90 nm, $h_\textrm{HSQ}$ = 70 nm with respect to Fig.~\ref{fig_1}a (i). \textbf{(b)} Measured and \textbf{(c)} simulated spectral dependence of light reflection for three different crystallization levels of L$^{0\%}$ (i.e., a-GST), L$^{40\%}$ (stands for $40\%$ crystallization level), and L$^{100\%}$ (i.e., c-GST) from the meta-switch. The experimental L$^{0\%}$, L$^{40\%}$, and L$^{100\%}$ levels are achieved by heating GST at 145 $^{\circ}$C for 0, 5, and 10 minutes, respectively.
	\textbf{(d)} Distribution of electric and magnetic fields at the fundamental resonances shown with circle, square, and triangle in panel \textbf{(c)}. Numerically resolved instantaneous electric field distribution is represented as vectors (top row) and magnetic field distribution is shown as colormap (bottom row) following plane-wave excitation with (i, ii) L$^{0\%}$, (iii, iv) L$^{40\%}$, and (v, iv) L$^{100\%}$. \textbf{(e)} Schematic illustration of the experimental setup for IR reflectometry and back focal plane imaging. The lens K-space is inserted for the beam deflection measurements.}
	\label{fig_2}
\end{figure*}

The calculated reflection profile for the meta-atom in Fig.~\ref{fig_1}a (i) as a function of wavelength ($\lambda$) for different pitch sizes ($p$) is shown in Fig.~\ref{fig_1}e (i). This figure clearly shows a profound reflection dip at a wavelength that varies with $p$. We ascribe the excitation of this ``hybrid'' resonance mode to the interference of the SPP mode, which is induced due to the metasurface periodicity providing an additional momentum, with the electric dipole moment, which is induced by enhanced light-matter interaction in the corners of the highly polarizable Si nanodisk (see Fig.~\ref{fig_1}d (i)). 

For the square array of meta-atoms in Fig.~\ref{fig_1}b (i), the momentum matching condition between the SPP and the in-plane component of the wavevector of the incident light can be described by Bragg's coupling equation \cite{pors2013efficient}:

\begin{align}
\textbf{k}_{||}(\lambda) & ~\pm~i\textbf{G}_{x}~\pm~j\textbf{G}_{y}~=~\textbf{k}_{\textrm{SPP}}(\lambda) \label{equ_PSPP}, \\
& \mid\textbf{k}_\textrm{SPP}(\lambda)\mid~=~\mid\textbf{k}(\lambda)\mid~\textrm{n}_{\textrm{SPP}}, \nonumber
\end{align}
wherein, $\textbf{k}_{||}(\lambda)$ is the in-plane wavevector at the incident wavelength $\lambda$, integers $i$ and $j$ are the grating orders for the reciprocal lattice vectors $\textbf{G}_{x}$ and $\textbf{G}_{y}$ ($|{\textbf{G}_{x}}| = |{\textbf{G}_{y}}| = {2\pi}/{p}$), and n$_{\textrm{SPP}}$ is the effective index of the SPP mode. 
Considering normal illumination, grating coupling arises when $p = \lambda_\textrm{SPP} \sqrt{i^{2}+j^{2}}$, where $\lambda_\textrm{SPP}$ is the wavelength of the SPP mode. The blue dashed curve in Fig.~\ref{fig_1}e (i) illustrates the coupling of the incoming wave to the $(i, j) = (1, 0)$ SPP at the Au-GST interface. Full-wave simulation is carried out to calculate n$_\textrm{SPP}$ of the SPP mode supported by the structure shown in Fig.~\ref{fig_1}b (i) in the absence of the nanodisk array. 

The gray-scale-coded dotted line in Fig.~\ref{fig_1}e (i) displays the spectral evolution of the electric dipolar resonance of a 400 nm-wide nanodisk array given that in our simulation the Au substrate is replaced by a perfect electric conductor (PEC) to isolate the effect of the SPP. Since this is a non-propagating localized mode bounded to the Si nanodisk end-faces, a dispersionless behavior is expected. Having a close look, this mode experiences weak dispersion for larger fill factors (i.e., $r/p$) in Fig.~\ref{fig_1}a (i), where each individual nanodisk slightly couples to its neighbors, while this effect fairly vanishes for higher values of $p$ (at a fixed $r$). The gray color bar in Fig.~\ref{fig_1}e (i), which represents the normalized field enhancement in the middle of two adjacent meta-atoms, well justifies this effect. In this regard, the electric dipole moment is strong enough to destructively interfere with the SPP mode and diminish the reflection dip for the lower pitch sizes ($p$). However, for the larger $p$, the overall reflection is naturally inclined to the SPP response as the dominant mode of the structure. This can be justified through Eq.~\ref{equ_PSPP}, in which by increasing the $p$ (or decreasing the reciprocal lattice component), the reflection dip of the metasurface, due to the grating coupling of the incident light to the SPP mode, redshifts. 

As shown in Fig.~\ref{fig_1}e (ii), by transforming a-GST to c-GST, the overall reflection response of the metasurface in Fig.~\ref{fig_1}b (i) significantly changes. The broadened feature of the resonance in this spectral window is mainly due to the intrinsic loss of the c-GST. We attribute the resonant mechanism here to the excited SPP mode (represented by the blue dashed line Fig.~\ref{fig_1}e (ii)) that is fairly overshadowed by the MD mode (represented by the gray-scale-coded dotted line). This emanates from: i) the remarkable refractive index contrast between c-GST and Si that hinders the effective diffraction of the incident light by nanodisks into the c-GST layer, and ii) more importantly, the large complex permittivity of c-GST, as the host medium of polaritons, in direct contact with the plasmonic metal that leads to the weak excitation of SPP mode \cite{maier2007plasmonics}.

To justify the existence of the MD as the dominant mode of the metasurface in the c-GST, we calculate the reflection spectrum as a function of the Si nanodisk thickness (i.e., $h_{\textrm{Si}}$ in Fig.~\ref{fig_1}a (i)) while considering the Au back reflector as a PEC (see Supplementary Fig.~3a). Due to the high effective refractive index of c-GST, we can treat the combination of the Si nanodisk and the subjacent GST segment with an effective volume comparable to that of the nanodisk as a ``high-index pseudo-pillar'', which is the host of the MD. Such a strong dipole, which is oriented perpendicular to the polarization of the circulating electric field inside the pseudo-pillar, is induced by the displacement current loop governed by the electric field of the incident light \cite{staude2013tailoring}. This effect significantly depends on the retardation of the incident field along the propagation direction of light in the pseudo-pillar. When the effective wavelength of light inside the pseudo-pillar becomes comparable to its size, the MD is excited. As a result, $h_\textrm{Si}$ has a key role in the MD excitation where by increasing $h_\textrm{Si}$, the reflection dip and the normalized magnetic field enhancement first increase and then decrease (see Supplementary Fig.~3a at the operational wavelength of $\lambda = 1100$ nm). Similar results are achieved by exploiting the generalized method of image applied to the multipole expansion approach to study the scattered light from the reflective hybrid meta-atom \cite{li2018generalized}. These results imply that there exists an optimum height at which the polarization of the incident electric field becomes anti-parallel at the vertical ends of the pseudo-pillar, and thus, the magnetic field of the incident light is highly enhanced at the center of the pseudo-pillar. Given the optimum $h_{\textrm{Si}}$, the gray-scale-coded dotted line in Fig.~\ref{fig_1}e (ii) shows that for the lower range of $p$ values, the MD moment affords higher normalized field enhancements. Thanks to the constructive interference between the SPP and the MD in this region, the deepest reflection dip forms at $p = 490$ nm. To get a deeper insight into the origination of fundamental modes, we carry out angle-resolved reflection calculation and parametric study in Supplementary Note~1 and Supplementary Fig.~4.  

Motivated by the rich underlying nature of the light-matter interaction unlocked by analytical calculations and full-wave numerical simulations, the remainder of this paper will be focused on the design, fabrication, and characterization of two on-demand classes of metasurfaces to control the amplitude and phase of the incident light. 

\subsection{Dynamic hybrid metasurface to control the amplitude response}

To show the unique capabilities of the metasurface architecture in Fig.~\ref{fig_1}, we design and experimentally demonstrate a tunable meta-switch with high modulation depth in a wide near-IR spectral range. 

The dynamic hybrid metasurface shown in Fig.~\ref{fig_1}b is fabricated in an area of 50 $\mu$m $\times$ 50 $\mu$m using a standard fabrication procedure including evaporation, thin-film deposition, lithography, and etching techniques (as described in Methods and Supplementary Fig.~2). Due to the material sensitivity of PCMs, some of the processes are customized to prevent possible degradation of GST during the fabrication. Figure~\ref{fig_2}a displays a scanning electron micrograph (SEM) of the as-fabricated tunable meta-switch. The meta-switch was optically characterized using an IR reflectometry measurement setup shown in Fig.~\ref{fig_2}e (see Methods for operation details). The measured and simulated reflection spectra at normal incidence for the two extremes and one intermediate state of the meta-switch are shown in Figs.~\ref{fig_2}b and \ref{fig_2}c, respectively, (see Supplementary Fig.~5 for more intermediate states). When the GST later is in its amorphous and crystalline phases, the hybrid metasurface shows a reflection dip of $\approx 0.19$ and $\approx 0.10$ located at $\lambda = 1515$ nm and $\lambda = 1260$ nm, respectively. Given that GST in its amorphous state has the lowest extinction coefficient, the dominant effect of SPP, associated with higher energy localization in the vicinity of the Au substrate, is to provide the required loss for a deep reflection in the C-band (1530 nm $\leq \lambda \leq$ 1565 nm). The optimum structural parameters enable the excitation of the MD mode that relatively sustains the resonance shape of the meta-atom with GST in the crystalline state. Due to the monolithic increase in the refractive index of GST from the amorphous to the fully-crystalline state, a continuous redshift of the resonance wavelength and thus, reflection dip in Fig.~\ref{fig_2}b is naturally expected. Regarding this, starting at L$^{0\%}$ (amorphous), the low-order resonance dip in Fig.~\ref{fig_2}b first progressively moves from the C-band to the L-band (1565 nm $\leq \lambda \leq$ 1625 nm) while its Q becomes gradually larger, then around L$^{60\%}$ the high-order hybridized mode emerges and smoothly redshifts until finally settling at the O-band (1260 nm $\leq \lambda \leq$ 1360 nm) for L$^{100\%}$ (fully-crystalline) (see Supplementary Fig.~5 for more details).

As depicted in Figs.~\ref{fig_2}b and \ref{fig_2}c, the experimental far-field reflection responses of the hybrid meta-switch under normal illumination are in good agreement with the simulated results. The redshift between the reflection dips of the experimental and simulated spectra stems from the fabrication imperfections, angular sensitivity to the inclined excitation, partial crystallization of GST during the fabrication process, finite size of the array of meta-atoms, and discrepancy between the optical constant of fabricated materials and those used in simulations. Moreover, the cross-sectional SEMs reveal that upon conversion from amorphous to fully-crystalline, the physical thickness of layer uniformly shrinks by $\approx 5\%$ (see Supplementary Fig.~6). 

The relative absorption bandwidth of the meta-switch, defined as $\textrm{BW}_{S(\lambda)}~=~2\times(\lambda_{l}-\lambda_{s})/(\lambda_{l}+\lambda_{s})$, in which $\lambda_{l}$ and $\lambda_{s}$ are the long and short limits of the wavelength range with absorption above 70\%, respectively, reaches about 1.3\%, 1.9\%, and 14.4\% for $LL{0\%}$, L$^{40\%}$, and L$^{100\%}$, respectively. Moreover, the relative modulation contrasts in reflection ($|{r_\textrm{a-GST} - r_\textrm{c-GST}}| /~\textrm{Max}~(r_\textrm{a-GST},~r_\textrm{c-GST})$), with $r_\textrm{a-GST}$ and $r_\textrm{c-GST}$ being the reflectivity of the structure with a-GST and c-GST, respectively, and the function Max ($a, b$) returning the larger of $a$ and $b$ are $0.56$ and $0.79$ at $\lambda = 1550$ nm and $\lambda = 1300$ nm, respectively.
Assuming $Q_{a}$ and $Q_{r}$ representing the metasurface resonance quality factors due to all intrinsic (non-radiation) and radiation losses, respectively, the meta-switch exhibits under-coupled (i.e., $Q_{a} < Q_{r}$) operation at the fundamental mode in the amorphous state and gradually shifts to the over-coupled regime (i.e., $Q_{a} > Q_{r}$) upon partial-crystallization of the GST layer and finally returns to the under-coupled regime upon being fully-crystallized (see Supplementary Note 2).

To understand the physical mechanism of light-matter interaction in each state, full-wave simulations are performed using a commercialized electromagnetic solver (see Methods for details). The electric and magnetic near-field distributions in a cut plane perpendicular to the surface of the meta-atom at the operational wavelength associated with the reflection dips in Fig.~\ref{fig_2}c are calculated to discern the origin of hybridized modes.

For the case of a-GST, two modes coexist in the hybrid meta-atom. The dominant one is SPP that originates from the collective oscillation of free electrons in the Au substrate coupled to the scattered light (from the 2D array of Si nanodisks in Fig.~\ref{fig_2}b) with an appropriate in-plane wavevector. This mode is distinguished by a mirrored electric field that semi-circulates perpendicularly to the Au surface (see Fig.~\ref{fig_2}d (i) and Fig.~\ref{fig_1}d (i)) and a magnetic field that oscillates in the center in parallel to the substrate surface (see Fig.~\ref{fig_2}d (ii) and Fig.~\ref{fig_1}d (i)). The interference between the launched SPPs along the positive and negative x directions forms an in-plane standing wave (see Supplementary Fig.~7a). The less influential mode for the a-GST case is the localized electric dipole mode associated with the high field intensity around the Si nanodisk edges (see Fig.~\ref{fig_2}d (i) and Fig.~\ref{fig_1}d (i)) and characterized by an enhanced electric field distribution at the lateral tips of the nanodisk (see Supplementary Fig.~7b). It is worth to notice that due to the coupling between two neighboring nanodisks, the electric field is remarkably enhanced (over 5.8$\times$) in the gap between the two meta-atoms. 

\begin{figure*}
	\centering
	\includegraphics[trim={0cm 0cm 0cm 0cm},width=0.92\textwidth, clip]{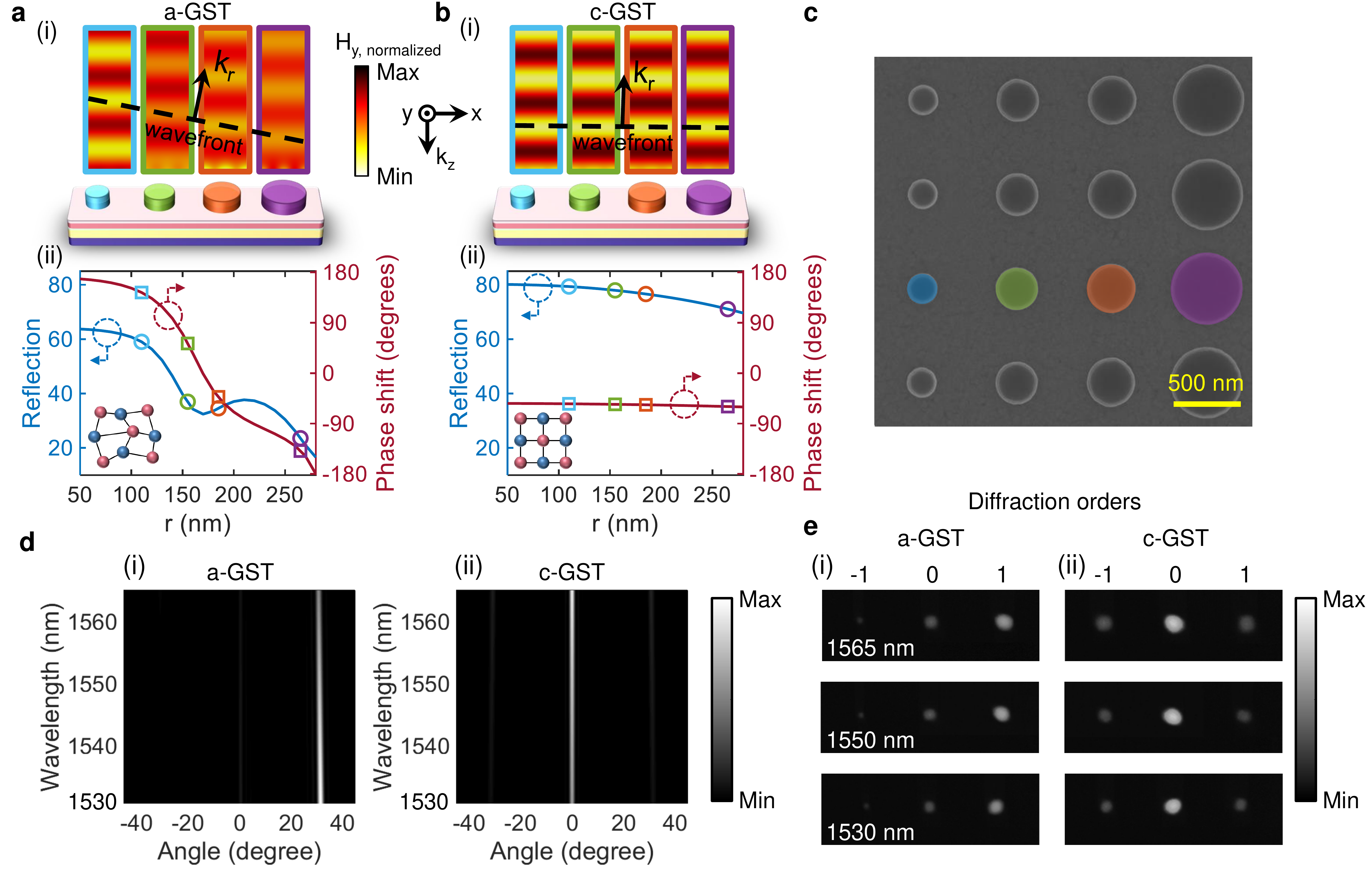}
	\caption{\textbf{Experimental realization of the dynamic meta-deflector}.
	\textbf{(a, b)} Wavefront evolution from a meta-molecule of the meta-deflector in Fig.~\ref{fig_1}c (i) in the case of \textbf{(a)} a-GST and \textbf{(b)} c-GST. (i) Scattered magnetic fields (captured at the same time instant) from the four constitutive meta-atoms of a meta-molecule. The phase shift increment induced by adjacent meta-atoms can change the wave peak position up to one wavelength (as shown by the black dashed line) in the amorphous state. Each strip is numerically calculated by an independent full-wave simulation of the corresponding meta-atom illuminated by a normally x-polarized plane wave at $\lambda = 1550$ nm. (ii) Evolution of reflectivity (left axis) and phase shift (right axis) of the meta-atom in \textbf{(a)} (see Fig.~\ref{fig_1}c (i)) as functions of the Si nanodisk radii. Discrete amplitude and phase values associated with each color-coded meta-atom in (i) are represented by circles and squares (inset: conceptual schematic of GST molecules). \textbf{(c)} Top view SEM of the as-fabricated sample. Si nanodisks associated with those sketched in Fig. \ref{fig_phase}a (i) are false-colored. \textbf{(d)} Simulated 2D maps for angular and spectral responses of the meta-deflector indicating anomalous reflection for a-GST (i) and specular reflection for c-GST (ii).
	\textbf{(e)} Measured normalized far-field reflection response of the meta-deflector obtained from the CCD camera in Fig.~\ref{fig_2}e for three different incident wavelengths covering the C-band.
	}
	\label{fig_phase}
\end{figure*}

By transforming GST to its fully-crystalline state (L$^{100\%}$), the effective wavelength of light inside the high-index pseudo-pillar becomes comparable to its size. This guarantees the excitation of the MD mode characterized by the circular displacement current of the electric field within the pseudo-pillar. Figure ~\ref{fig_2}d (v) (and Fig.~\ref{fig_1}d (ii)) shows that the orientation of the electric field is antiparallel at the top of the Si nanodisk and the bottom of the GST layer, which is associated with maximized magnetic field at the center of the pseudo-pillar (as shown in Fig.~\ref{fig_2}d (vi)). Furthermore, the resonating horizontal MD mode at the bottom end of the pseudo-pillar results in the formation of a localized plasmonic hotspot at the surface interface of the Au substrate and the GST film, as depicted in Fig.~\ref{fig_2}d (vi). Such an enhanced mode couples the surface-normal propagating plane wave to the highly attenuating SPP (see Fig.~\ref{fig_1}d (ii) for a more clear view).

The physical mechanism of governing modes are also investigated through full-wave simulations at resonant wavelengths associated with four intermediate states of GST. Numerical results show that the metasurface in Fig.~\ref{fig_2}a demonstrates a fairly similar behavior for GST states between amorphous (L$^{0\%}$) and semi-crystalline states up to L$^{60\%}$, and thereafter the governing modes follow distributions relatively similar to that of the fully-crystalline state (L$^{100\%}$). For the sake of brevity, the electric and magnetic fields for the case of L$^{40\%}$ are shown in Figs.~\ref{fig_2}d (iii) and \ref{fig_2}d (iv), and the remaining results are provided in Supplementary Fig.~5.

\subsection{Dynamic hybrid metasurface to control the phase response}

The governing plasmonic-photonic modes of the metasurface in Fig.~\ref{fig_1} can enable dynamic phase control and wavefront engineering in a rather wide wavelength range. To prove this unique feature, the meta-atom in Fig.~\ref{fig_1}a (i) is re-optimized to simultaneously provide wide (narrow) phase span with maximized scattering efficiency in the amorphous (crystalline) state at an operation wavelength around $\lambda = 1550$ nm. The meta-atom cross-section remains circular for polarization-insensitive operation, and the structural parameters are $h_{\textrm{HSQ}} = 90$ nm, $h_{\textrm{Si}} = 100$ nm, $h_{\textrm{SiO}_{2}} = 5$ nm, $h_{\textrm{GST}} = 70$ nm, and $p = 720$ nm. As shown in Fig.~\ref{fig_phase}a, by varying the Si nanodisk radius (i.e., $r$), almost full $360^{\circ}$ optical phase coverage is achieved in the amorphous state, which is remarkable for any phase-based optical functionality. This phase profile then follows a flat trace upon transformation of a-GST to c-GST (see Fig.~\ref{fig_phase}b). It is notable that the reflection amplitude response of the metasurface in the a-GST case experiences considerable variation with $r$ due to the on-resonance operation of the metasurface, as shown in Fig.~\ref{fig_phase}a (ii). This situation is significantly relaxed for the c-GST case due to the off-resonance operation associated with fairly flat amplitude response with a higher overall optical efficiency (see Fig.~\ref{fig_phase}b (ii)).

To realize dynamic beam deflection, we exploit the generalized Snell's law defined as \cite{yu2011light}:
\begin{equation}
{\theta_{r}} = \textrm{sin}^{-1}(\dfrac{\lambda\Delta\phi}{2\pi{p}} + \textrm{sin}{\theta_{i}}),
\end{equation}
in which $\theta_{r}$ and $\theta_{i}$ are the reflection and incidence angles, respectively, and $\Delta\phi$ represents the phase increment between the adjacent meta-atoms (see Fig.~\ref{fig_phase}a). We leverage a 2D array of meta-molecules comprised of four different meta-atoms introducing a linear phase gradient into the incoming wavefront to scatter light to the anomalous (specular) angle for a-GST (c-GST) case. For the case of a-GST, $\Delta\phi \approx 90^{\circ}$ is chosen associated with $r_{1} = 110$ nm, $r_{2} = 155$ nm, $r_{3} = 185$ nm, and $r_{4} = 265$ nm to facilitate constructive interference at the outgoing angle of $\theta_{r} \approx 32^{\circ}$ under normal illumination (see Fig.~\ref{fig_phase}a (i)). By converting a-GST to the c-GST, all designed meta-atoms serve as similar phase retarders (i.e., $\Delta\phi \approx 0^{\circ}$). As a result, the overall optical response of the meta-deflector resembles that of a mirror-like metasurface that reflects incident light to the specular angle (i.e., $\theta_{r} = \theta_{i}$), as shown in Fig.~\ref{fig_phase}b (i). Although the extinction coefficient of c-GST is large at $\lambda = 1550$ nm, its high complex refractive index effectively makes the device mismatched to the vacuum impedance while reducing the time of light-matter interaction, and thus alleviates the total loss. 


\begin{figure*}
	\centering
	\includegraphics[trim={0cm 0cm 0cm 0cm},width=0.9\textwidth, clip]{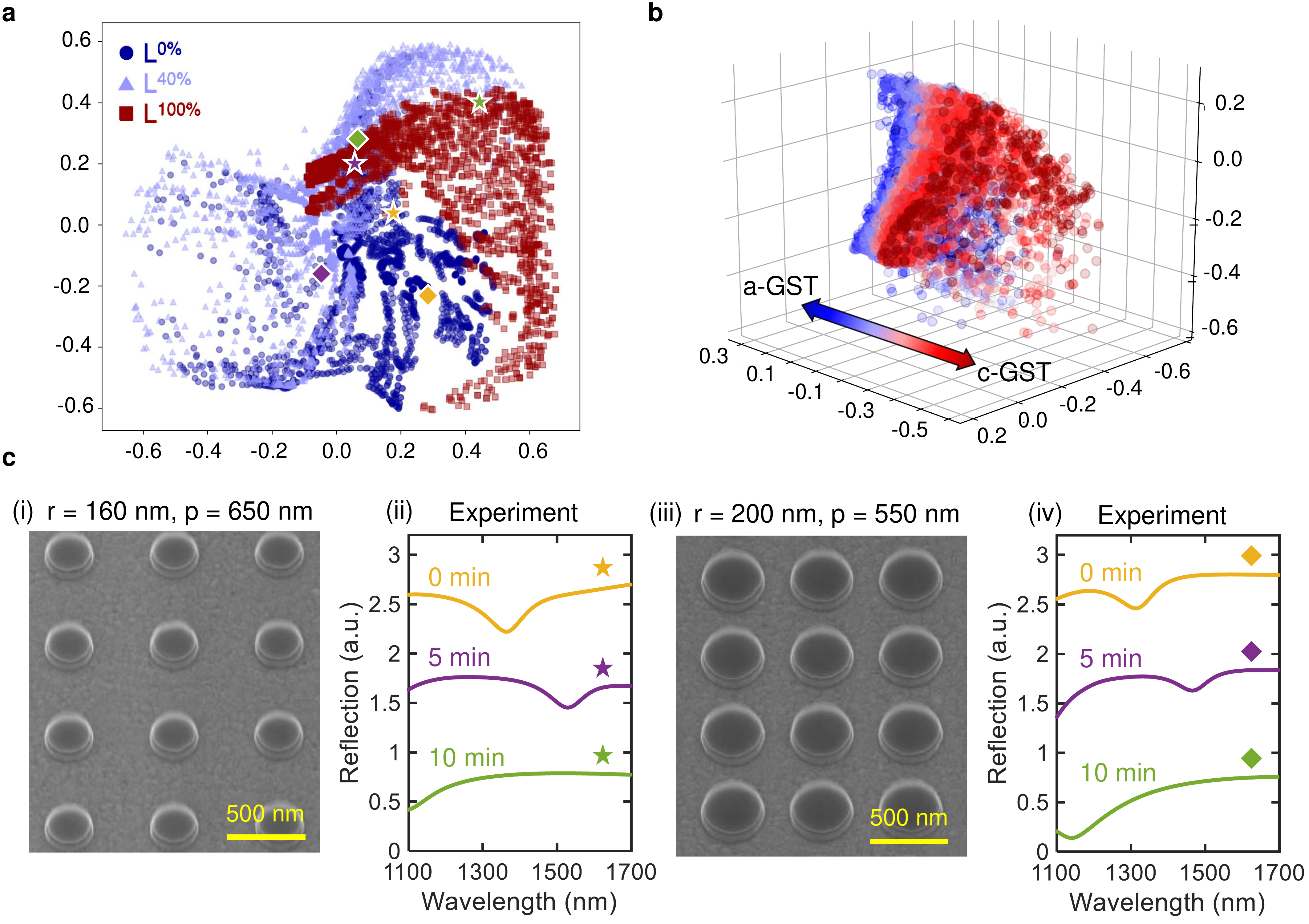}
	\caption{\textbf{Low-dimensional visual representation of simulated reflection responses using manifold learning}. 
	\textbf{(a)} 2D manifolds of reflection responses with GST in three crystallization levels (i.e., L$^{0\%}$, L$^{40\%}$, and L$^{100\%}$). 
	\textbf{(b)} 3D manifolds of reflection responses with GST in 11 equally distributed crystallization levels from amorphous to the fully-crystalline states.
	\textbf{(c)} SEMs (i, iii) and measured reflection spectra (ii, iv) for two sets of metasurfaces with different $r$ and $p$ and same geometrical parameters of $h_{\textrm{HSQ}} = 90$ nm, $h_{\textrm{Si} = 100}$ nm, $h_{\textrm{SiO}_{2}} = 5$~nm, and $h_{\textrm{GST}} = 70$ nm. The 2D representation of each spectrum is shown with a solid marker in panel \textbf{(a)}. Orange, purple, and green curves represent reflection spectra for L$^{0\%}$, L$^{40\%}$, and L$^{100\%}$, respectively.}
	\label{LLE_2and3D}
\end{figure*}

The PCM-functionalized meta-deflector is fabricated (see the SEM in Fig.~\ref{fig_phase}c) following the established techniques explained in Methods and characterized using a back focal plane imaging setup as detailed in Fig.~\ref{fig_2}e. Figures~\ref{fig_phase}d and \ref{fig_phase}e demonstrate the 2D maps for the simulated and experimentally measured far-field angular-spectral reflection response of the  meta-deflector under normal excitation for the a-GST and c-GST cases.
Figure~\ref{fig_phase}e shows that in the amorphous state $\theta_{r} \approx 30^{\circ}$ at three different incident wavelengths in the C-band, which is in good agreement with the simulation results. Moreover, the overall efficiency slightly differs from the numerical simulation results. By transforming GST to the crystalline state, the meta-deflector reflects the incident light with an ordinary specular angle of $\theta_{r} \approx 0^{\circ}$ and an efficiency marginally lower than that predicted by simulations due to the excitation of symmetric first-order diffractions. In both cases, the fabrication imperfections (surface roughness, deviation from a designed shape/height, partial crystallization of GST, etc), higher material losses of the fabricated sample than those of the modeled materials, oxidation of the GST layer during the transfer process between different chambers, limited extension of the array of meta-molecules, and light leakage from sample edges are the main sources of discrepancy between simulation and experimental results. Further sources can likely be the finite range of incident/collected angles and light coupling to higher order modes originating from non-perfect normal excitation of the sample.

\subsection{Low-dimensional representation of response reconfiguration using manifold learning}

To demonstrate the ability of the dynamic hybrid metasurface in Fig.~\ref{fig_1} in providing a large range of responses (beyond those of a single static metasurface) through GST phase conversion, we leverage a nonlinear dimensionality reduction approach \cite{hinton2006reducing}, namely manifold learning. This technique provides an intuitive visual representation of the detailed reflection spectra of the hybrid metasurfaces with different design properties and GST phases in a low-dimensional space. It also enables to follow the evolution of the metasurface responses upon changes in the GST phase or the geometrical parameters in the meta-atom structure. The meta-atom in Fig.~\ref{fig_1}a (i) has six geometrical design parameters, which are $r$, $p$, $h_{\textrm{HSQ}}$, $h_{\textrm{Si}}$, $h_{\textrm{SiO}_{2}}$, and $h_{\textrm{GST}}$. To generate the training data for the manifold learning algorithm, we calculate the refection spectra of 2400 different meta-atoms with geometrical parameters selected randomly. For each set of geometrical parameters, GST with 11 equi-spaced crystallization levels (L$^{0\%}$, L$^{10\%}$, L$^{20\%}$, ..., L$^{100\%}$) are simulated resulting in total 26400 simulations. Each of the calculated reflection spectra is uniformly sampled at 1000 wavelengths in the desired wavelength range of 1100-1700 nm, resulting in a 1000-dimensional response space for the meta-atom. To find the low-dimensional visual representation of the reflection spectra, we apply the autoencoder algorithm \cite{liu2018generative,kiarashinejad2019deep, kiarashinejad2020deep, liu2020compounding,kiarashinejad2020knowledge} to the 26400 refection responses while keeping the mean-squared error as a measure of the accuracy of the algorithm. In parallel, we train an autoencoder using 7200 responses calculated for L$^{0\%}$, L$^{40\%}$, and L$^{100\%}$ structures. Figure \ref{LLE_2and3D}a shows the loci of the responses in the 2D reduced space (or the latent space) for these three GST crystallization levels. Each manifold (identified by a distinct color) represents the range of responses available from the metasurface with one of the three GST crystalline levels. It is clear that by changing the GST phase, a large range of possible responses in the latent space is covered, which is considerably larger than that available from a single structure. Figure \ref{LLE_2and3D}b shows the responses for all 11 crystallization levels of GST in a three-dimensional (3D) latent space demonstrating the capability of the dynamic metasurfaces in covering a far larger range of responses than that available for a single (static) structure. In addition, Fig.~\ref{LLE_2and3D}b shows the gradual evolution of the reflection spectral response through changing the GST crystallization level. Figures \ref{LLE_2and3D}a and \ref{LLE_2and3D}b clearly show the advantage of using the reconfigurable metasurfaces in providing a large range of programmable responses for a metasurface enabled by the large dynamic range of the GST optical properties at different crystallization levels. 

To corroborate this observation, we fabricate two sets of metasurfaces with different $r$ and $p$ as shown in Figs.~\ref{LLE_2and3D}c (i) and \ref{LLE_2and3D}c (iii). The corresponding measured refection spectra for the L$^{0\%}$, L$^{40\%}$, and L$^{100\%}$ cases of these two structures are displayed in Figs.~\ref{LLE_2and3D}c (ii) and \ref{LLE_2and3D}c (iv), respectively. The dimensionality-reduced experimental results are also shown in Fig. \ref{LLE_2and3D}a. While all these figures show the wide range of tuning of the reflection dip through GST phase change, Fig.~\ref{LLE_2and3D}a clearly shows that the experimental response for each structure with a distinct crystallization level belongs to its theoretically calculated manifold in the latent space. 

It is important to note that the large range of responses for the reconfigurable metasurfaces in this paper is achieved by having only one blanket layer of GST. It is imagined that by using multiple independent GST patches a considerably larger response range from a metasurfaces might be possible. This will be a major step towards creation of all-purpose spatial light modulators with unprecedented level of control over spatial, spectral, and even temporal properties of an optical wavefront.

\section{discussion}

A key enabling feature of the metasurfaces discussed in this paper is the possibility of engineering the meta-atom with a range of electromagnetic modes through interplay between SPP, MD, and electric dipole. By integrating with PCMs and utilizing the huge dynamic range of their optical properties through crystalline phase change, the hybrid platform in this paper provides all major requirements for practical metasurfaces with subwavelength reconfigurable unit cells. We believe that further fundamental studies on complete engineering of the spatial, spectral, and polarization properties of the electromagnetic modes of the meta-atoms and combining them to form meta-molecules can enable a systematic design approach for forming large-scale metasurfaces for several state-of-the-art applications such as imaging, spectroscopy, computing, and quantum photonics. 

Besides the wealthy physics governing the operation of associated reconfigurable hybrid metasurfaces, the demonstrated platform can address some existing technological challenges with available devices. The low thermal time constant of the system (due to the plain plasmonic metal substrate) can facilitate the reversible phase switching of GST via Joule heating (especially during melt-quenching process). This is in contrast to most of previous demonstrations in which the PCM is incorporated into the plasmonic meta-atoms (with a melting temperature comparable to that needed to excite the phase-change process of the PCM) that are vulnerable to deformation in the melt-quenching process needed for the re-amorphization of PCMs. Moreover, the high electrical resistivity and thermal conductivity of the double-purpose metallic inclusion can facilitate in-situ Joule heating of the realized meta-device, which leads to fully integrable device platforms necessary for practical scenarios. Furthermore, the highly confined plasmonic mode with pronounced filed intensity in the vicinity of the metallic part, lowers the required energy for transformation of GST between its two extreme states.

While our results show the value of adding PCMs to metasurfaces for reconfiguration, further research is needed to bring reconfiguration to the subwavelength dimensions. Our results are achieved by using a blanket PCM layer for the entire metasurface. The next step is to fabricate structures with 2D PCM patches that can be independently controlled. It is also essential to use electrical PCM phase control by using proper wiring using transparent electrical wiring (e.g., using ITO \cite{taghinejad2020ito}). Fortunately, there is a lot of knowledge from the field of data storage in independent addressing of miniaturized GST regions, which could be adopted for reconfigurable metasurfaces. This will be a major advantage over alternative architectures that have major challenges to address 2D pixels with good dynamic range and spatial resolution. Finally, the platform discussed here has superior performance in terms of speed and power consumption. PCMs like GST can operate with 10's-100's ns reconfiguration times \cite{wuttig2017phase}. Their non-volatile phase-change mechanism also removes any DC power consumption, and the reconfiguration power consumption is due to the control signals during the phase-change process. As an example, energies in the order of 19.2 aJ/nm$^{3}$ (6.6 aJ/nm$^{3}$) for crystallization (amorphization) of GST have been reported recently \cite{zheng2020modeling}. This is acceptable for many practical applications.

In summary, the metasurface architecture demonstrated here provides a unique platform for practical applications through fundamental mode engineering and subwavelength reconfiguration with an unprecedented dynamic range. It can motivate new research and development both at the fundamental-science level and the device/system-design level.

\section{Funding}

The presented work was supported primarily by the Office of Naval Research (ONR)(N00014-18-1-2055, Dr. B.
Bennett), and in part by Defense Advanced Research Projects Agency (DARPA) (D19AC00001, Dr. M. Fiddy).

\section{Acknowledgments}

This work was performed in part at the Georgia Tech Institute for Electronics and Nanotechnology (IEN), a member of the National Nanotechnology Coordinated Infrastructure (NNCI), which is supported by the National Science Foundation (NSF) (Grant ECCS1542174). 

\section{Methods}

\subsection{Numerical simulations}

All numerical simulations are carried out by using the commercial software packages CST Microwave Studio based on the finite integral technique and COMSOL Multiphysics based on the finite element method. For the design of meta-atoms, periodic boundary conditions in the x and y directions of the unit cell are employed. Also, the perfectly matched layer (PML) is considered in the z direction to mimic a free-space to monitor far-field scattering (see Supplementary Fig.~8). Simulations are performed in a 3D computational domain using a non-uniform mesh topology with hexahedral elements in all directions. The maximum element size of $\lambda$/10, where $\lambda$ corresponds to the shortest wavelength in the analyzed spectral window, is chosen. A plane wave is launched into the meta-atom along the z direction, and the far-field reflection spectrum is monitored at the input port. Electric and magnetic field distributions are detected within the field profile monitors. Optical properties of all materials are obtained from ellipsometry measurements (see Supplementary Fig.~1) except for the Au layer which is described by the Lorentz-Drude model with three times damping constant larger than the bulk material \cite{johnson1972optical}. Amongst the existing effective-medium theories, we use the Lorentz-Lorenz relation to model the effective dielectric constants of GST in different crystallization levels as follows: 

\begin{equation} 
\frac{\epsilon_{eff}(\lambda)-1}{\epsilon_{eff}(\lambda)+2}=m\times\frac{\epsilon_{c}(\lambda)-1}{\epsilon_{c}(\lambda)+2}+(1-m)\times\frac{\epsilon_{a}(\lambda)-1}{\epsilon_{a}(\lambda)+2},
\label{Lorentz-Lorenz}
\end{equation} 
where for a specific wavelength $(\lambda)$, $\epsilon_{c}(\lambda)$ and $\epsilon_{a}(\lambda)$ are the permittivities of crystalline and amorphous GST, respectively, and ${m}$, ranging from 0 (associated with L$^{0\%}$) to 1 (associated with L$^{100\%}$), is the crystallization level of GST. The optical properties of GST in the intermediate states are reflected in Supplementary Fig.~1.

\subsection{Sample fabrication}

To keep the intrinsic optical properties of the as-deposited GST film intact, an exhaustive optimization of the fabrication procedure is carried out. First, a prime Si substrate (500 $\mu$m-thick) is cleaned in acetone within an ultrasound bath; rinsed using methanol, isopropyl alcohol, and deionized water; dried using dust-free nitrogen, and exposed to the oxygen plasma. A successive deposition of a titanium (Ti) adhesion layer (5 nm-thick, 0.2 A s$^{-1}$ at 2$\times$10$^{-6}$ mbar) and an Au film (100 nm-thick, 0.3 A s$^{-1}$ at 2$\times$10$^{-6}$ mbar) is then carried out using an electron-beam evaporation system without breaking the vacuum. Then, the sample is placed in the atomic layer deposition (ALD) chamber to thermally grow a SiO$_{2}$ film (5 nm-thick at 150$~^{\circ}\textrm{C}$) using a standard two-pulse process of water and TDMAS precursors (20 s and 60 s, respectively). In the next step, the sample is transformed to the sputtering chamber for RF magnetron deposition of a layer of GST (70 nm-thick, background pressure of 1$\times$10$^{-6}$ mbar and 45 W power under 40 sccm argon (Ar) flow). It is quickly followed by another ALD deposition process to form the second SiO$_{2}$ film (5 nm-thick at 90 $^{\circ}$C) as a protective layer. Hydrogenated amorphous Si (a-Si) (90 nm-thick) is then deposited using plasma-enhanced chemical vapor deposition (PECVD) (with a 5$\%$ mixture of silane in Ar) at low temperature to prevent crystallization of as-deposited GST. Negative electron-beam resist (hydrogen silsesquioxane (HSQ) 6$\%$) is then spin-coated (60 s at 6000 r.p.m.) and baked (180 s at 90 $^{\circ}$C). Then, a water-soluble anti-charging conductive polymer is spin-coated (Espacer 300Z, 30 s at 2000 r.p.m.) to avoid static charging in the patterning process. Electron-beam lithography (EBL) is performed with 100 kV acceleration voltage, 120 $\mu$m aperture, 1 nA beam current, 6.4 nm exposure step size, 2500 $\mu$C.cm$^{-2}$ dose, and 500$\times$500 $\mu$m$^{2}$ writing field in an Elionix ELS-G100 system. In the next step, the anti-charging layer is removed in a water bath, and the exposed pattern is developed in a resist developer (TMAH for 30 s at 40$~^\circ{\textrm{C}}$) followed by deionized-water rinsing (5 min) and nitrogen drying. The etching process is then performed using chlorine (Cl$_{2}$) gas in an inductively coupled plasma reactive ion etcher (ICP-RIE), in which a detailed optimization of the etching process is conducted to minimize the exposure of the GST layer to the plasma while maximizing the sidewall sharpness. The left-over HSQ (with the final optimized thickness) on the Si nanodisks is not removed at the end. An  explanatory  figure  showing  the  fabrication  steps can be found in Supplementary Fig.~2. Also, we elaborate on such a stacked configuration in Supplementary Note~4.


\subsection{Thermal and optical switching experiments}
To transform the phase of GST from amorphous to semi- and fully-crystalline, the sample under test is placed at the center of a wide hotplate with a fixed temperature of 145 $^{\circ}$C. The prescribed time for 40\% and 100\% crystallization levels are 5 and 10 minutes, respectively. After being heated, the sample is cooled down in the ambient, then transferred to the optical setup for the progressive reflection measurement. We also demonstrate sub-micron-size crystallization of a metasurface-less GST layer using a train of ultrashort laser pulses (using Nanoscribe GmbH) as shown in Supplementary Fig.~9. To highlight the potential of GST as a reversible functional material, we employ a commercialized femtosecond laser setup (Optec WS-Flex) to amorphize a spot size of $15~\mu$m on a thermally crystallized metasurface-less GST layer. The exact composition and thickness of the GST film as well as the thermal conductivity of the medium in contact with the GST layer defines the fluence of the ultrashort pulse for a uniform conversion. In our experiment, we find that a single 6 nJ$\mu\textrm{m}^{-2}$ pulse can successfully amorphize a 70-nm-thick c-GST layer.

\subsection{Optical characterization}

The optical measurements are preformed in a homemade reflectometry setup. The filtered collimated beam from a tungsten light source is coupled to an optical fiber to illuminate the $50~\times~50~\mu$m$^{2}$ pattern using a 20$\times$ objective lens with numerical aperture (NA) of 0.4. The reflected light is collected and focused by a tube lens, a spectrometer, and a silicon CCD (charge-coupled device) camera. All the wavelength-dependent data collected by the spectrometer are normalized with respect to that of a bare silver mirror in the measurements of the meta-switch. By introducing and removing a lens K-space near the CCD (see Fig. \ref{fig_2}e), we can image the back focal plane of the objective lens and switch between the real space and the Fourier plane. The bright field images are taken using a reflection microscopy setup illuminated by a white light source and equipped with a digital CCD camera. All measurements are carried out at room temperature (25 $^\circ{ \textrm{C}}$).

\subsection{Material characterization}

The complex refractive index of GST thin films is determined using a Woollam Ellipsometer m2000. The spectroscopic ellipsometery measurements are performed at three angles of incidence (50$^{\circ}$, 60$^{\circ}$, 70$^{\circ}$) over a spectral range from 350 nm to 1750 nm. Oscillator parameters as well as the thickness of the GST film and the surface roughness are used as fitting parameters. Tauc-Lorentz and Cody-Lorentz models are
employed for the evaluation of optical functions of thin GST films in as-deposited and crystalline states \cite{shportko2008resonant}. The model parameters including Lorentz oscillator amplitude, resonance energy, oscillator width, optical bandgap, and Urbach energy are chosen as fitting parameters. The surface roughness measured by atomic force microscopy (AFM) measurements in both phases is fixed in all calculations.

Supplementary Figure~9a illustrates the 3D surface topography images using  AFM for as-deposited and fully-crystalline (using a uniform thermal process) thin films of GST. The roughness of the GST film associated with the number of fine grains fairly increases by converting a-GST to c-GST. Supplementary Figure~9b shows the 2D surface image of a line ($\approx~1~\mu$m-wide) written with a train of optical laser pulses on a thin film of GST. The thickness of the optically phase-converted GST region appears $\approx 3.5$ nm ($\approx 5\%$ out of a 70 nm-thick GST film) lower than that of the a-GST film due to an increase in the film density by the crystallization process \cite{li2016reversible}. The results are in good agreement with the measurements of thermally crystallized GST films.

To determine the binding energies of the core electrons in a-GST films, X-ray photoelectron spectroscopy (XPS) is performed. To remove surface contamination as well as oxidation, the sample is sputtered with Ar ions with an energy of 0.5 KeV and a current density of approximately 10 $\mu$A.cm$^{-2}$ resulting in a hole size of tens of micrometers. This leads to the improvement of the intensity peaks of the main elements in the survey scan of the alloy. We carry out the survey scan within the binding energy range of 0-600 eV for different etched thicknesses, and core-level spectra of elements are plotted in Supplementary Fig.~10. 

We use an inVia Qontor confocal Raman microscope (100$\times$ objective) to study the Raman scattering in the micro-Raman configuration from a thin film of GST in as-deposited, crystalline, and re-amorphized states illuminated by the primary 785 nm laser. The power and integration time are set to 0.3 mW and 10 s, respectively, in all cases to prevent unwanted crystallization and possible ablation during measurements. Supplementary Figure~11 clearly shows that the as-deposited (and amorphous) state poses a rather broad peak, which converts to a dual-band peak upon transition to the crystalline phase. These results are in good agreement with the $\textit{ab initio}$ molecular dynamic simulations and experimental results available in the literature \cite{kolobov2004understanding}. The sole peak in the amorphous state is due to vibrations of defective octahedra formed by Te, Sb, and a majority of the Ge atoms while the induced dual peak in the crystalline state corresponds to the larger spread in the Ge-Te and Sb-Te bond lengths \cite{sosso2011raman}. 

PANalytical Empyrean diffractometer is employed to study the X-ray diffraction (XRD) spectrum of the as-deposited and fully-crystalline states of a thin film of GST. The corresponding XRD patterns of a 170 nm-thick layer of GST deposited on a Si wafer are depicted in Supplementary Fig~12, in which the corresponding curves are shifted along the vertical axis for the sake of clarity. The XRD spectrum displayed for the crystalline state shows Bragg peaks justifying the NaCl type structure with a face-centered cubic configuration \cite{park2009optical}.



\providecommand{\noopsort}[1]{}\providecommand{\singleletter}[1]{#1}%

\clearpage
\newpage



\end{document}